\documentclass[12pt]{article}
\usepackage{epsf,latexsym,indentfirst}
\input{epsf}

\def\be{\begin{equation}}
\def\ee{\end{equation}}
\def\bea{\begin{eqnarray}}
\def\eea{\end{eqnarray}}
\def\bx{{\bf x}}
\def\hal{{1\over 2}}

\author{
F.Karsch$^1$\footnote{karsch@Physik.Uni-Bielefeld.DE}\,,
M.Oevers$^{1,2}$\footnote{oevers@Physik.Uni-Bielefeld.DE}\,,
P. Petreczky$^{1,3}$\footnote{petreczk@Physik.Uni-Bielefeld.DE\vskip0truecm
\hspace{0.28cm}petr@hercules.elte.hu}\,\,,\\
$^1$Fakult\"at f\"ur Physik, Universit\"at Bielefeld,\\
\qquad  P.O. Box 100131, D-33501 Bielefeld, Germany\\
\makebox[0.5cm]{}$^2$Department of Physics and Astronomy,\\ 
University of Glasgow, Glasgow,  G12 8QQ, U.K.\\
\makebox[0.5cm]{}$^3$Dept. of Atomic Physics, E\"otv\"os University,\\
H-1088 Puskin 5-7, Budapest Hungary \\
}
\title{
\small
\begin{flushright}
BI-TP 98/19 \\
ZIF-MS-33/98 \\
revised, September 1998
\end{flushright}
\vspace {1.5cm}
\Large  \bf Screening Masses of Hot 
SU(2) Gauge Theory\\
from the 3d Adjoint Higgs Model
\vskip0.5truecm
}
\normalsize
\date{\vskip1.5cm}

\begin{document}
\maketitle
\thispagestyle{empty}
\begin{abstract}
\noindent
We study the Landau gauge propagators of the lattice 
$SU(2)$ 3d adjoint Higgs
model, considered as an effective theory of high temperature 4d $SU(2)$ gauge
theory. From the long distance behaviour of the propagators we extract
the screening masses. The propagators are studied both in the symmetric
and the broken phases of the 3d Higgs model. It is shown that the pole
masses extracted from the propagators in the symmetric phase agree 
well with the screening masses obtained recently in finite
temperature $SU(2)$ theory, while
propagators measured in the broken phase show quite a different
behaviour. 
This suggest that the symmetric phase of the
3d model corresponds
to the deconfined phase of the 4d $SU(2)$ gauge theory. The relation of the
propagator masses to the masses extracted from gauge invariant
correlators and the mass gap of pure 3d $SU(2)$ gauge theory is also
discussed.
\end{abstract}

\newpage
\pagenumbering{arabic}

\section{Introduction}
The screening of static chromo-electric fields is  one of the most outstanding
properties of $QCD$ and its investigation is important both from
a theoretical and phenomenological point of view (for phenomenological
applications see e.g. \cite{wang}).  In leading order of perturbation
theory the associated inverse screening length (Debye mass) is defined
as the $IR$ limit of the longitudinal part of the gluon self energy
$\Pi (k_0=0,{\bf k} \rightarrow 0)$. However, as the screening
phenomenon is related to the long distance behaviour of $QCD$ 
the naive perturbative definition of the Debye mass
is obstructed by severe $IR$ divergences of
thermal field theory and beyond leading order the above definition
is 
no longer
applicable. It was suggested by Rebhan \cite{rebhan} to define
the Debye mass as a pole of the longitudinal part of the gluon
propagator. Such a definition implies a self-consistent resummation of the
perturbative series and ensures the gauge-independence and $IR$ finiteness of
the Debye mass. However, it requires the introduction of a so-called
magnetic screening mass, a concept introduced long ago \cite{linde}
to cure the $IR$ singularities of finite temperature non-Abelian theories.
Analogously to the electric (Debye) mass the magnetic mass can be
defined as a pole of the transverse part of the finite temperature 
gluon propagator. Though the magnetic mass is not calculable in 
perturbation theory different self-consistent resummation schemes of the
perturbative series give a non-vanishing magnetic mass \cite{eberlein,
jackiw, alexanian, buchmuller1,buchmuller11}. The non-zero magnetic mass is also
clearly seen in lattice studies of propagators in $SU(2)$
gauge theory \cite{heller1,heller2}.

As the screening masses are static quantities it is expected that they
can be determined in
a 3d effective theory of $QCD$, the 3d $SU(3)$ adjoint Higgs model, provided
the temperature is high enough. However, in the case of $QCD$ one may
worry whether the standard arguments of  dimensional reduction
apply. First of all the coupling constant is large $g \sim 1$  
for any physically interesting temperature and thus the requirement
$g T << \pi T$ is not really satisfied. Moreover, it is at present not
clear  which phase of the effective theory corresponds to
the high temperature phase of $QCD$. The 3d adjoint Higgs model is known
to have two phases, the symmetric (confinement) phase and the Higgs
(Coulomb) phase \cite{nadkarni, hart1, kajantie1}. The perturbative
calculation of the effective potential \cite{kajantie1,polonyi} in the 
effective theory 
suggests
that the Higgs phase corresponds to the deconfined
phase of 4d $SU(3)$ theory. This conclusion seems to be supported
by the 2-loop level dimensional reduction performed in \cite{kajantie1}.
On the other hand in the dimensional reduction approach applied
in \cite{lacock1,karkkainen2,karkkainen1} the symmetric phase turns out
to be the physical one and a good description of spatial Wilson loops and
Polyakov loop correlators has been obtained.

The aim of the present paper is to clarify whether the screening masses,
defined as  poles of the corresponding lattice propagators in 
Landau gauge, can be determined in the effective theory for the simplest
case of the $SU(2)$ gauge group, where precise 4d data on  screening
masses  exist for a huge temperature range \cite{heller2}.
The generalization to the $SU(3)$ case and the inclusion of 
fermions\footnote{Fermionic
fields do not appear in the effective theory, their role is only to 
modify the parameters of the effective theory.} 
is then straightforward.

There are several reasons which make the study of the screening masses,
defined as poles of the corresponding propagators, in the effective
theory interesting.
First of all 
it is possible to compare masses defined in this way directly with 4d
measurements and thus clarify the status of the 3d adjoint Higgs model as 
an effective theory. 
Secondly  the masses defined in this way are  closest to
the spirit of resummation techniques, which implicitly rely on
dimensional reduction (see e.g. discussion in \cite{patkos1}). 
A third reason is that gauge invariant definitions of the Debye mass rely
on the 3d effective theory \cite{kajantie1,arnold,kajantie2}
and although these definitions are the best
in the sense that they are explicitly gauge invariant, the corresponding
correlators have not
been measured yet  in the 4d theory. 
Gauge invariant definitions yield larger masses than the pole masses
\cite{kajantie1, kajantie2}
and further studies are required to establish the connection between them.
We 
will try to relate these masses in the 
last section of our paper in the spirit of the so-called constituent model
\cite{buchmuller2}.

\section{The 3d SU(2) Adjoint Higgs Model on Lattice}
The lattice action for the 3d adjoint Higgs model used in the present paper
has the form
\bea
&&
S=\beta \sum_P \hal Tr U_P + 
\beta \sum_{\bx,\hat i} \hal Tr A_0(\bx) U_i(\bx) A_0(\bx+\hat i)
U_i^{\dagger}(\bx) + \nonumber\\
&&
\sum_{\bx} \left[-\beta\left(3+\hal h\right) \hal Tr A_0^2(\bx) + 
\beta x { \left( \hal Tr
A_0^2(\bx)\right)}^2 \right],
\label{act}
\eea
where $U_P$ is the plaquette, $U_i$ are the usual link variables and
the adjoint Higgs field is parameterized, as in \cite{kajantie1,
lacock1,karkkainen1} by anti-hermitian matrices $A_0=i \sum_a \sigma^a
A_0^a$ ($\sigma^a$ {are the usual Pauli matricies}). Furthermore 
 $\beta$ is the lattice gauge coupling, $x$ parameterizes the 
quartic self coupling of the Higgs field and $h$ denotes the
bare Higgs mass squared. In principle the  parameters appearing in
eq.~(\ref{act}) can be related to the parameters of the original 4d theory
via the procedure of dimensional reduction
\cite{kajantie1,lacock1,karkkainen2,karkkainen1}, 
which is essentially perturbative.
Another possible procedure is  to find the above parameters by
matching some quantities which are equally well calculable both in the full
4d  lattice theory and in the effective 3d lattice theory, 
i.e. to perform a non-perturbative matching.
Since the validity of the perturbative dimensional reduction, as was
discussed before, is not obvious in the case of $QCD$ (or more 
precisely for $SU(N_c)$ 
gauge theories)  we will try to explore the 
latter
approach in the present paper.
In general this would require a matching analysis in a 3d parameter space
($\beta, ~x,~h$), which is clearly difficult in general. We thus followed
a more moderate approach and fix two of three parameters
namely $\beta$
and $x$, to the values obtained from the perturbative 
procedure of dimensional reduction. The values of these parameters
at 2-loop level are \cite{kajantie1}
\vskip0.3truecm
\bea
&&
\beta={4\over g_3^2 a}, \nonumber \\[0.1cm]
&&
g_3^2=g^2\left(\mu\right) T \left[1+{g^2\left(\mu\right)\over 16 \pi^2}
\left(L+{2\over3}\right)\right], \\
&&
x={g^2\left(\mu\right)\over 3 \pi^2}\left[1+
{g^2\left(\mu\right)\over 16 \pi^2} \left(L+4\right)\right],\\
&&
L={44\over 3} \ln{\mu\over 7.0555 T},
\eea
with $a$ 
and $T$ denoting the lattice spacing and temperature, respectively.
The coupling constant of the 4d theory $g^2(\mu)$ is defined through 
the 2-loop formula
\be
g^{-2}(\mu)={11\over 12 \pi^2} \ln{\mu\over \Lambda_{\overline{MS}}}+
{17\over 44 \pi^2}\ln\left[2\ln{\mu\over
\Lambda_{\overline{MS}}}\right].
\label{4dg}
\ee
In order to be able to compare the results of the 3d simulation with 
the corresponding ones
in the 4d theory it is necessary to fix the renormalization and 
the temperature
scale. We choose the renormalization scale to be $\mu=2 \pi T$, which ensures
that corrections to the leading order results for the parameters
$g_3^2$ and $x$ 
of the effective theory are small. Furthermore we use the 
relation $T_c=1.06 \Lambda_{\overline{MS}}$ from \cite{heller2}. 
Now the temperature scale is fixed 
completely and the physical temperature may be varied by varying the 
parameter $x$.  The lattice spacing was chosen according to the criterium
$a<<m^{-1}<<N a$, where $N$ is the extension of the lattice and $m$ is the
mass we want to measure. 

The main goal of the present investigation is to study the propagators  of
scalar and vector (gauge) fields. For this purpose one has 
to fix a specific gauge,
which is chosen to be the Landau gauge
$\partial_{\mu} A_{\mu}=0$ \footnote{the choice of this gauge is
motivated by the fact that 4d Landau gauge condition
used in \cite{heller1,heller2} for static field configuration is
equivalent to the 3d Landau gauge condition}. On the lattice this is realized by
maximizing the quantity:
\be
Tr \left[ \sum_{\bx,i} \left ( U_i(\bx)+U^{\dagger}_i(\bx) \right) \right]
\label{max}
\ee
The gauge fixing is performed using the overrelaxation algorithm, which 
in our case is as efficient as combined overrelaxation and $FFT$ algorithm 
used in \cite{heller1,heller2}.
The vector field is defined in terms of link variables as
\be
A_i(\bx)={1\over 2 i} (U_i(\bx)-U^{\dagger}_i(\bx))
\ee
We are interested in extracting the electric (Debye) $m_D$ and the magnetic
$m_T$ screening masses from the long distance behaviour of the scalar and
vector propagators defined as
\bea
&&
G_D(z)=< Tr A_0(z) A_0^{\dagger}(0) > \sim \exp(- m_D z), \\
&&
G_T(z)= {1\over 2} ( G_1(z) + G_2(z)) \sim \exp(- m_T z),
\eea
with  
\bea
&&
G_i(z)=< Tr A_i(z) A_i(0) >, \nonumber\\
&&
A_\mu(z)=\sum_{x,y} A_\mu(x,y,z),\quad \mu=0,~1,~2
\eea
Note that due to the Landau gauge condition $G_3(z)$ should be constant. This
fact can be used to test the precision and validity of the gauge fixing 
procedure. In our case this condition is satisfied with an accuracy of $0.1\%$.

Besides the scalar and vector propagators of the adjoint Higgs model 
we also calculate the
gauge invariant scalar correlators and analyze the propagators in the limit
of a 3d pure gauge theory.

To extract the masses from the correlation functions we have used the
general fitting ansatz
\be
A \left[ {\exp\left(-m z\right) \over {z}^b}+ {\exp \left(- m \left (N_z-z
\right)\right)\over {\left(N_z-z\right)}^b} \right]\quad  
\label{fitI}
\ee
motivated in \cite{bernard1} through the analysis of the gauge dependence
of the electron propagator in QED.
Previous investigations of gauge boson and quark propagators in Landau gauge
\cite{heller1,bernard1,dim} have shown that effective masses extracted from
the correlation functions rise with increasing Euclidean time separation
and eventually reach a plateau. This is contrary to local masses extracted 
from gauge invariant correlation functions which approach a plateau from 
above and is a direct consequence of a non-positive transfer matrix in 
Landau gauge. A fit with $b\ne 0$ can account for the observed drift in
local masses and allows to extract stable masses already at shorter distances.  

Most of our numerical studies have been performed on lattices of size\footnote{We
also have performed additional calculations on a
$16^2\times 32$ lattice and checked explicitly that results for the electric
mass show no volume dependence. Furthermore some calculations have
been performed on a $32^2\times 96$ lattice to check that the lattice size
used for our calculations was sufficient for the determination of the
magnetic mass.} $32^2\times 64$. 
From previous studies of gluon propagators in the 4-dimensional SU(2)
gauge theory we know that such large lattices are needed to observe a plateau in
local masses \cite{heller2}. This is in particular the case for 
the rather small magnetic screening mass which leads to a rather slow decay
of the correlation functions. We thus use the same spatial lattice size as in 
those studies.

We have used correlated (Michael-McKerrell) fits with eigenvalue smearing
\cite{michael}. 
Our fits have been constrained to the region where local masses show a 
plateau (typically $z\sim 15$ for the magnetic mass and $z\sim 5$ for 
the electric mass). In this region fits with $b=0$ and $b\ne 0$ yield 
consistent results within statistical errors. From fits with $b\ne 0$ we 
find best fits with $b < 0$, in accordance with the 
behaviour of local masses discussed above. However, 
although these fits with $b <0$ can start at shorter distances, $z$, and still
yield a good $\chi^2/d.o.f$, the magnitude of $b$ is not well determined
within our present statistical accuracy. In the following we thus will
quote results from fits with $b=0$.

\section{ Numerical Results for the Propagators of the 3d $SU(2)$ 
Higgs Model}

The deconfined high temperature 
phase of the 4d $SU(2)$ field theory corresponds to some
surface 
$h=h(x,\beta)$ in the parameter space ($\beta, x, h$) of the adjoint Higgs
model. This is the surface of 4d physics and 
may lie in the symmetric or the broken phase.
For fixed value of $\beta$ (lattice spacing) the 4d physics is 
described by a line on this surface. Since most of our simulation were done
for $\beta=16$ we will refer later to this line as the line of 4d
physics.
In  the present paper three choices for the line of 4d physics
$h(x)$ are explored. A comparison with 4d simulations should allow to
determine the physical line $h(x)$ which reproduces the result of the
4d analysis.
The first choice for $h(x)$ 
is the perturbative line of 4d physics, {calculated in \cite{kajantie1}
and} lying in the broken phase,
the other two choices are in the symmetric phase.
The three choices for the line of 4d physics are illustrated on Figure 1 for
$\beta=16$.
The actual procedure we used to  choose the parameter $h$ in the
symmetric phase is the following.
First we have determined the transition line $h_{tr}(x)$. The transition line
as function of $x$ in the infinite volume limit was found in
\cite{kajantie1} in terms of the renormalized mass parameter
$y=m^2/g_3^4$ ($m$ is the continuum renormalized mass). The transition
line in terms of $y$ turns out to be independent of $\beta$.
Then using eq. (5.7) from \cite{kajantie1} one can calculate $h_{tr}(x)$.
\begin{figure}
\vspace{-1.5cm}
\epsfysize=10cm
\epsfxsize=12cm
\centerline{\epsffile{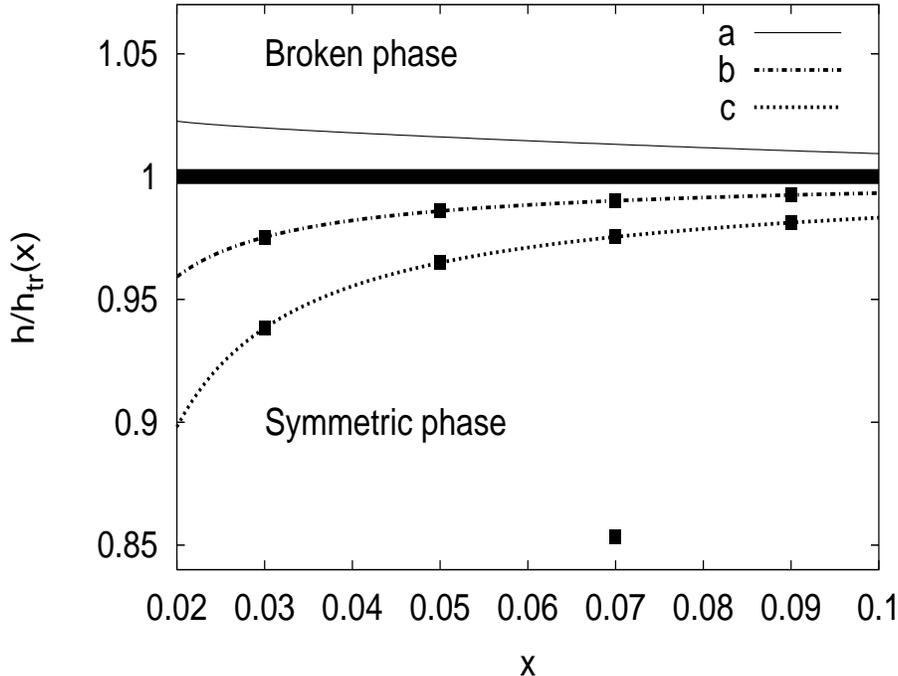}}
\vspace{-1cm}
\caption{The bare masses normalized by the critical mass $h_{tr}(x)$ 
used in our matching analysis (squares): perturbative line (a) and
two sets $I$ (line b) and $II$ (line c). For a discussion of their
choice see text. For $x=0.07$ also a point deeper in the symmetric phase
has been selected.
The thick solid line is the transition line and the
thickness of the line indicate the  uncertainty in its value.}
\end{figure}
The usage of the infinite volume result for the transition
line seems to be justified 
because most of our simulation were done  on a
$32^2 \times 64$ lattice. 
The two sets of $h(x)$ values, which appear on Figure 1,
were chosen so that the renormalized mass
parameter $y$ (calculated using eq. (5.7) of \cite{kajantie1} ) always 
stays $10 \%$ and $25 \%$ away from the transition line.
These values of $h$ are of course {\em ad hoc} and one should use them only as
trial values.
The values of the parameters in the symmetric phase are shown 
in Table 1 where the two sets of $h$ values are 
denoted as (I) and (II) and also the values
of $h_{tr}$ corresponding to the transition line are given.
\vskip0.5truecm
\begin{center}
\begin{tabular}{|l|l|l|}
\hline
$Temperature~scale~~$ &$~~~~~~~~~~~~~~~~~~~~~~~~~~~~~~h~~~~~~~~~~~~~~~~~~~~~~~~~~~~~$ \\
\end{tabular}
\begin{tabular}{|l|l|l|l|l|}
\hline 
$~~~~x~~~~$  &$~~~~T/T_c~~~~$   &$~~~~~~~I~~~~~~~$  &$~~~~~~~II~~~~~~~$ 
&$~~~~transition~~~~$\\
\hline
$~~0.09~~$  &$~~~4.433$   &$~~-0.2652$   &$~~-0.2622$   &$~~-0.2672(4)$\\
$~~0.07~~$  &$~~~12.57$   &$~~-0.2528$   &$~~-0.2490$   &$~~-0.2553(5)$\\
$~~0.05~~$  &$~~~86.36$   &$~~-0.2365$   &$~~-0.2314$   &$~~-0.2399(6)$\\
$~~0.03~~$  &$~~~8761$   &$~~-0.2085$   &$~~-0.2006$   &$~~-0.2138(9)$\\
\hline
\end{tabular}
\vskip0.3truecm
Table 1: {\em The two sets of the
bare mass squared used in the simulation and those
which correspond to the transition line for $\beta=16$}
\end{center}

\begin{figure}
\vspace{-1cm}
\epsfysize=9cm
\epsfxsize=12cm
\centerline{\epsffile{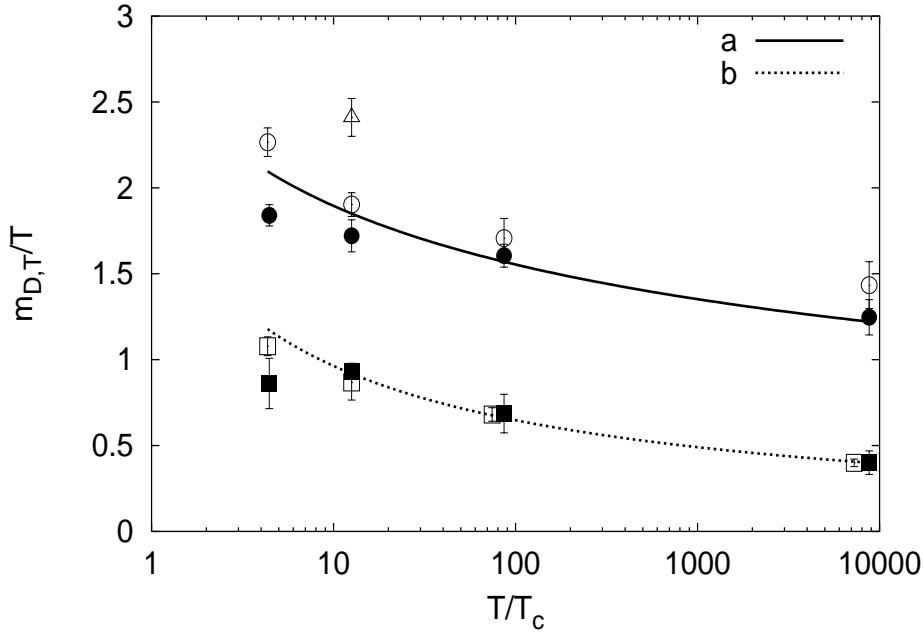}}
\vspace{-0.7cm}
\caption{The screening masses in units of the temperature. 
Shown are the Debye mass
$m_D$ for the first (filled circles) and the second (open circles)
set of $h$, and
the magnetic mass  $m_T$ for the first (filled squares)
and the second (open squares) set of $h$.
The line (a) and line (b) represent the fit for the temperature
dependence of the Debye and the magnetic mass from 4d
simulations
from \cite{heller2}. The magnetic mass for the set $II$
at the temperature $T \sim 90
T_c$ and $ \sim 9000 T_c$ was shifted in the temperature scale for
better visualization. The open triangle is the value of the Debye mass
for $x=0.07$ and $h=-0.2179$}
\end{figure}

Let us first discuss our calculations in the broken phase.
In the broken phase the simulations were done for two sets of parameters:
$\beta=16,~~x=0.03,~~h=-0.2181$ and $\beta=8~~,x=0.09,~~h=-0.5159$,
here $h$ was chosen along the perturbative line of 4d physics,
which has been calculated in \cite{kajantie1} to 2-loop order.
The propagators obtained by us in the broken phase show 
a behaviour which is very different from
that in the symmetric phase and that in the 4d case studied in Refs.
\cite{heller1, heller2}. The magnetic mass extracted from the 
gauge field propagators is $0.104(20)g_3^2$ for the first set of 
parameters and $0.094(8)g_3^2$ for the second set of parameters.
It thus is a factor 4 to 5 smaller than the corresponding 4d result.
Moreover, the propagator of the $A_0$ field does not seem to
show a simple exponential behaviour, this fact actually is in qualitative 
agreement with findings of Ref. \cite{rebhan}.
Taken together these facts suggest that the broken phase does not
correspond to the physical phase.

Let us now turn to a discussion of our results in the symmetric phase.
In order to find the parameter range of interest for $h$ we first
have analyzed three different values of $h$ at $\beta=16$ and $x=0.07$.
In addition to the two values given in Table~1 we also chose an even
larger value, $h=-0.2179$. The location of these values relative to 
the transition line is shown in Figure~1. For the electric screening
mass we find, for increasing values of $h$,
$m_D/T = 1.72(10)$, $1.90(7)$ and $2.41(11)$. 
We note that $m_D/T$ increases with 
increasing distance from the transition line. These results should be 
compared to the 4d data. From the fit given in Ref.\cite{heller2}, we find at 
$T/T_c= 12.57$ for the electric screening mass  $m_D/T = 1.85$. This 
shows that our third value for $h$ clearly is inconsistent with the 
4d result. One has to choose values for $h$ close to  the transition
line in order to get agreement between the 3d and 4d results. 
From a linear interpolation between the results at the three different
values for $h$ we find the best matching value, i.e. a point on the 
line of 4d physics, $h(x=0.07)=-0.2496$. 

The analysis described above motivated our choice of trial values for
$h$ at other values of $x$ as given in Table~1.
The temperature dependence of the screening masses 
obtained in the symmetric phase for these two sets of  
parameters, which stay close to the transition line, is shown in Figure~2. 
Also given there is the result of the 4d simulations \cite{heller2}, $m_D^2/T^2 = Ag^2(T)$,
with $A=1.70(2)$ for the electric mass and $m_T/T= C g^2(T)$, with
$C=0.456(6)$ for the magnetic mass. As can be seen both masses can
be described consistently with a single choice of the coupling $h$ for
temperatures larger than $10T_c$. Although even at $T\simeq 4T_c$ we 
find reasonable agreement with the 4d fits, we note that the dependence
of the results on the correct choice of $h$ becomes stronger and a 
simultaneous matching of the electric and magnetic masses seems to be
difficult. For larger temperatures we find that the magnetic mass 
shows little dependence on $h$ (in the narrow range we have analyzed) 
and the determination of the correct choice of $h$ thus is mainly
controlled by the variation of the electric mass with $h$.  

Let us summarize our findings for the screening masses in the
symmetric phase.  For $T \ge 10 T_c$ the screening masses can be
described very well in the effective theory, provided that values of $h$
are close to the transition line. The suitable values of $h$ can be
found using the interpolation procedure outlined above for $x=0.07$. 
This procedure can be also done for $x=0.05$ and $0.03$, however there
the 4d data are well described by values corresponding to the set
$I$ (see Figure 2), therefore the following $h$ values can be considered
as ones corresponding to 4d physics, $h(x=0.07)=-0.2496$, 
$h(x=0.05)=-0.2365$ and $h(x=0.03)=-0.2085$. An interpolation between
these values gives the line of 4d physics $h_{4d}(x)$.

Before closing this section we briefly want to address the question of a 
possible gauge dependence of the screening masses extracted by us.
The pole of the propagator was proven to be gauge invariant in
perturbation theory \cite{kobes1,kobes2,kronfeld}. As concerns the
results on the propagator pole masses extracted from lattice simulation 
the situation is less clear. Here one faces the numerical problem
to isolate the asymptotic large distance behaviour of the correlation
function from possible short distance (powerlike) corrections which are
gauge dependent \cite{bernard1}.  
In \cite{dim} quark-propagators in axial, Coulomb and Landau
gauges were studied and the effective masses extracted at quite short
physical distances from the 
propagators were found to depend on gauge. In \cite{bernard1} the
quark and gluon propagators were investigated in so-called
$\lambda$-gauges. Masses extracted from the propagators using
exponential fits like Eq.~(\ref{fitI}) also show a dependence (although
mild) on the gauge parameter.
To study the gauge dependence of our results following \cite{bernard1}
we have introduced $\lambda$-gauges which in our case are defined by the
condition
\be
\lambda \partial_3 A_3+\partial_2 A_2+\partial_1 A_1=0.
\ee
and measured the propagators on $32^2 \times 96$ lattice. The masses were
extracted from the propagators using the functional form given in 
Eq.~(\ref{fitI}).
It turns out that screening masses measured at present for one parameter set
and $\lambda=0.5, 1.0$ and 2.0 are the same within our statistical 
errors of about (5-10)\%. Such an accuracy is sufficient for the matching
analysis and the resulting conclusions given above.
Nonetheless, a more detailed analysis of the gauge dependence of the
propagator pole masses is certainly needed. Work on this is in progress and
will be presented elsewhere.

\section{Magnetic Mass in 3d $SU(2)$ Pure Gauge Theory}

The magnetic mass found in the previous section seems to scale with
the 3d gauge coupling $g_3^2$. This behaviour can be understood in the
following way: If the temperature is high enough, the separation of
different length scales holds, i.e. $g^2 T<<g T<< 2 \pi T$ and besides
non-static modes with mass $\sim 2 \pi T$ the heavy $A_0$ field  with mass
$\sim g T$ can also be integrated out. In this limit the $IR$ behaviour of
high temperature 
{\bf $SU(N_c)$ }
gauge theory is described by 3d pure gauge theory in
which the only mass scale is $g_3^2$.
Unfortunately for $QCD$ (or $SU(2)$ gauge theory) the above arguments fail to 
hold because the coupling remains large for any realistic temperature. 
A non-perturbative study is therefore needed to establish the 
relation between
the magnetic mass found in finite temperature $SU(2)$ theory and the mass gap 
of 3d pure gauge theory.

We have measured the Landau gauge propagators for the 3d $SU(2)$ gauge theory
and from its large distance behaviour extracted the magnetic mass.
The results for different values of $\beta$ are listed in Table~2.
\vskip0.3truecm
\begin{center}
\begin{tabular}{|l|l|l|l|l|l|}
\hline
$~~~\beta~~~~$  & $~~~m_T/g_3^2~~~$ & $~~~\chi^2/d.o.f~~~$\\
\hline
$12.00$ & $0.48\pm0.036$ & $~~~~~0.580 $\\
\hline
$16.00$ & $0.42\pm0.070$ & $~~~~~0.497 $\\
\hline
$20.00$ & $0.44\pm0.068$ & $~~~~~1.439 $\\
\hline
\end{tabular}
\vskip0.5truecm
Table 2: {\em The results of the fit for the magnetic mass in 3d pure
gauge theory}
\end{center}

Using the data from Table~2 one finds $m_T=0.46(3)g_3^2$.
This value  is in good agreement with  the 3d adjoint Higgs
model result. The magnetic mass thus  is rather insensitive to the dynamics
of the $A_0$ field. 
This finding is 
in accordance with the gap equation study of the 
adjoint Higgs model \cite{patkos1}.
It is instructive to consider also
the ratio of the magnetic mass and the string
tension. For $SU(2)$ pure gauge theory the latter was found in 
Ref. \cite{teper}
$\sqrt{\sigma_3}=0.334(3) g_3^2$, which yields 
$m_T/\sqrt{\sigma_3}=1.39(9)$. This should be compared with the ratio
of the magnetic mass and the spatial string tension of the high
temperature $SU(2)$ theory 
$m_T/\sqrt{\sigma_s}=1.27(1)(1+0.11(2) g^2(T))$ \cite{heller1}.

\section{Gauge Invariant Correlators and the Constituent Model}

Let us finally 
discuss the relation between gauge invariant and gauge dependent 
correlators.
Gauge invariant correlators for the $SU(2)$ Higgs model were studied in
detail in \cite{kajantie1}. 
We have studied here
the Polyakov loop correlator, which in the effective theory following
\cite{lacock1,karkkainen2} is defined as
\be
{<L_{eff}(\bx) L_{eff}(0)>\over {< \bar L_{eff} >}^2},~~~
L_{eff}(\bx)=\hal Tr \exp(A_0(\bx)), ~~~~~~
\bar L_{eff}=\sum_{\bx} L_{eff}(\bx)~~~~.
\ee
Furthermore we have analyzed
the correlation function of the scalar operator $Tr A_0^2$ 
whose large distance
behaviour  gives the mass of the $A_0-A_0$ bound state $m(A_0)$.
The mass of this
bound state is also expected to determine the exponential fall off of the 
Polyakov loop correlator \cite{kajantie1}. 
We have measured these correlators on a $16^2 \times 32$ lattice in the 
symmetric phase. As expected both  yield the same mass. The 
results for the masses extracted from the correlation function of $Tr
A_0^2$ are shown in Table~3. The values of these masses are consistent
with those obtained in \cite{kajantie1}.
For $x=0.09$, which corresponds to the temperature $T \sim 4 T_c$,
we find 
$m(A_0)/T=3.89(11)$ which should be compared with the Polyakov loop 
correlator in the high temperature 4d $SU(2)$ theory $m_P/T \sim 4$
\cite{engels}.
Let us discuss the relation of $m(A_0)$ and the Debye mass defined through
the propagator.
If the coupling constant is small enough then
$m(A_0) \sim 2 \sqrt{N\over 3} g T \equiv 2 m_{D0}$. In our case this 
perturbative relation is not 
satisfied. However, in terms of the constituent model \cite{buchmuller2}
the mass of the 
$A_0-A_0$ bound state can be represented by the sum of two constituent
scalar masses, which are $m_D$.  

The mass of the 
lowest lying glueball state was measured in
\cite{kajantie1}
and  was found to be rather independent of the
couplings $h$ and $x$ of the $A_0$ field, the value of this mass
is  $m_G \sim 2.0 g_3^2$.
In terms of the constituent model this state can  be viewed as a bound state 
of four constituent gluons \cite{buchmuller2} with a constituent mass equal 
to $m_T$, thus $m_G \sim 4 m_T$.
The masses predicted by the constituent model compared with the results
of direct measurements are shown in Table~3.
\vskip0.3truecm
\begin{center}
\begin{tabular}{|l|l|l|l|}
\hline
$~~~~parameters~~~~$  &$~~~~~~~~~~~~~~~~~m(A_0)/g_3^2~~~~~~~~~~~~$
&$~~~~~~~~~~~~~~~m_G/g_3^2~~~~~~~~~~~~~$\\
\hline
$\beta~~~\vline~~~~x~~~~\vline~~~~h~~~~~$ &$~measured~$ \vline $~constituent
~model~$ &$~measured$ \vline $~constituent~model~$\\
\hline
$16~~~~0.09~~~-0.2622$  &$~~~1.54(15)~~~~~~~~~~~~1.78(6)$
&$~~~~~~~~~~~~~~~~~~~~~~~~1.68(9)$\\
$16~~~~0.05~~~-0.2314$  &$~~~2.28(20)~~~~~~~~~~~~2.38(16)$
&$~~~2.0~~~~~~~~~~~~~~~~~1.88(17)$\\
$24~~~~0.03~~~-0.1475$  &$~~~3.03(65)~~~~~~~~~~~~3.28(30)$
&$~~~~~~~~~~~~~~~~~~~~~~~~1.84(10)$\\
\hline
\end{tabular}
\vskip0.3truecm
Table~3: {\em The masses of the scalar bound state measured on $16^2 \times
32$ lattice and the glueball mass taken from \cite{kajantie1} 
in units of $g_3^2$ compared with
the predictions of the constituent model}
\end{center}
\vskip0.3truecm
Another gauge invariant operator, whose correlation function  was
measured in \cite{kajantie1}, is $h_i=\epsilon_{ijk} Tr A_0 F_{jk}$.
The mass extracted from this correlator  is the mass of the bound state
of the scalar field and the light glue, thus its mass in terms of
the constituent model is expected to be $m_h=m_D+m_T$.
Our numerical simulations show  that $m_D>m_T$ 
in the entire temperature range, therefore one would expect that
$m(A_0)>m_h$. 
On the other hand the numerical simulation in Ref. \cite{kajantie1}
shows that this condition is satisfied only in the parameter range which
does not corresponds to the physical situation.
In this respect the constituent model seems to
fail to explain the spectrum of the theory.

\section{Conclusions}
In this paper we have investigated the propagators of the 3d
$SU(2)$ adjoint Higgs model in Landau gauge. Masses
extracted from them are compared with the corresponding ones from 
recent simulations of the high temperature $SU(2)$ theory \cite{heller2}.
The gauge coupling and the Higgs
quartic coupling  of the effective theory 
were fixed via the procedure of  dimensional reduction, while
the bare Higgs mass $h$ was left free to allow for an 
non-perturbative matching.
We have considered the values of $h$ which  correspond to the broken
phase and two sets of $h$ in the symmetric phase. The screening masses
in the symmetric
phase are in good agreement with the results of the 4d theory
with both set of $h$, which means that the propagator masses
in the symmetric phase are not too sensitive to the value of the bare mass.
In the broken phase the propagators are 
quite different from
those in the
symmetric phase and 4d simulation. In particular the scalar propagator
does not seem to show an exponential behaviour, while the vector 
propagators yield several times smaller masses than what is obtained in
4d simulation.
The magnetic mass in the adjoint Higgs model scales with $g_3^2$
and its value is very close to the
value of the mass gap obtained in the pure 3d gauge theory.
We have also measured gauge invariant correlators of the scalar field.
The resulting
mass as well the mass of the lightest glueball can be related to
the propagator masses via the constituent model proposed in 
\cite{buchmuller2}. The mass of the vector bound state,
measured in \cite{kajantie1},
however, 
cannot be understood in this way.

\medskip
\noindent
{\Large \bf Acknowledgements:}
This work was partially supported by the TMR Network {\em Finite Temperature
Phase Transition in Particle Physics}, EU contract no. ERBFMRX-CT97-0122 
and the ZiF project {\em Multi-scale
Phenomena on Massively Parallel Computers}.
We thank K. Rummukainen for providing us with data allowing to check
our program. 
P.P. thanks W. Buchm\"uller, Z. Fodor, A. Jakov\'ac, D. Miller and
A. Patk\'os for usefull discussions.

\end{document}